\documentclass[10pt,conference]{IEEEtran}
\IEEEoverridecommandlockouts

\usepackage[noadjust]{cite}
\usepackage{amsmath,amssymb,amsfonts}
\usepackage{algorithmic}
\usepackage{graphicx}
\usepackage{textcomp}
\usepackage{xcolor}
\usepackage[utf8]{inputenc}
\usepackage{newunicodechar,graphicx}
\usepackage[hidelinks,bookmarks=true]{hyperref}
\usepackage[numbered]{bookmark}
\usepackage{soul}
\usepackage{booktabs}
\usepackage{multirow}
\usepackage{float}
\usepackage{url}
\usepackage{tcolorbox}
\usepackage{tabularx} 
\usepackage[switch]{lineno}
\usepackage{enumitem}
\usepackage{seqsplit}
\usepackage{pifont}
\usepackage{balance}
\usepackage{microtype}
\usepackage{pgf-pie}  
\usepackage{fontawesome}

\hypersetup{
	pdfborder          = {0 0 0},
	breaklinks         = true,
	bookmarksnumbered  = true,
	pdfsubject         = {6th International Conference on Technical Debt (TechDebt 2023)},
	pdfkeywords        = {Software Quality, Refactoring, Technical Debt, Self-Admitted Technical Debt, SATD, TD, TechDebt, Android, Empirical Study},
	pdftitle           = {An Exploratory  Study on the Occurrence of Self-Admitted Technical Debt in Android Apps},
	pdfauthor          = {Gregory Wilder II,
Riley Miyamoto, Samuel Watson, Rick Kazman, Anthony Peruma}
}
\usepackage{listings}

\definecolor{javared}{rgb}{0.6,0,0} 
\definecolor{javagreen}{rgb}{0.25,0.5,0.35} 
\definecolor{javapurple}{rgb}{0.5,0,0.35} 
\definecolor{javadocblue}{rgb}{0.25,0.35,0.75} 
 
\DeclareRobustCommand{\okina}{%
  \raisebox{\dimexpr\fontcharht\font`A-\height}{%
    \scalebox{0.8}{`}%
  }%
}
\newunicodechar{ʻ}{\okina}

\newcommand{\RQA}{\textbf{RQ1}: To what extent do open-source Android app developers document technical debt in the code? }
\newcommand{\RQB}{\textbf{RQ2}: What are the specific types of self-admitted technical debt occurring in open-source Android apps?}
\newcommand{\RQC}{\textbf{RQ3}: What code elements in Android apps are susceptible to technical debt?}

\begin{document}

\title{An Exploratory Study on the Occurrence of Self-Admitted Technical Debt in Android Apps}



\makeatletter
\newcommand{\linebreakand}{%
  \end{@IEEEauthorhalign}
  \hfill\mbox{}\par
  \mbox{}\hfill\begin{@IEEEauthorhalign}
}
\makeatother

\author{\IEEEauthorblockN{Gregory Wilder II}
\IEEEauthorblockA{
\textit{University of Hawaiʻi at Mānoa}\\
Honolulu, Hawaiʻi, USA\\
gwilder@hawaii.edu}
\and
\IEEEauthorblockN{Riley Miyamoto}
\IEEEauthorblockA{
\textit{University of Hawaiʻi at Mānoa}\\
Honolulu, Hawaiʻi, USA\\
rkam9@hawaii.edu}
\and
\IEEEauthorblockN{Samuel Watson}
\IEEEauthorblockA{
\textit{University of Hawaiʻi at Mānoa}\\
Honolulu, Hawaiʻi, USA\\
scwatson@hawaii.edu}
\linebreakand
\IEEEauthorblockN{Rick Kazman}
\IEEEauthorblockA{
\textit{University of Hawaiʻi at Mānoa}\\
Honolulu, Hawaiʻi, USA\\
kazman@hawaii.edu}
\and
\IEEEauthorblockN{Anthony Peruma}
\IEEEauthorblockA{
\textit{University of Hawaiʻi at Mānoa}\\
Honolulu, Hawaiʻi, USA\\
peruma@hawaii.edu}
}

\maketitle

\begin{abstract}
Technical debt describes situations where developers write less-than-optimal code to meet project milestones. However, this debt accumulation often results in future developer effort to live with or fix these quality issues. To better manage this debt, developers may document their sub-optimal code as comments in the code (i.e., self-admitted technical debt or SATD). While prior research has investigated the occurrence and characteristics of SATD, this research has primarily focused on non-mobile systems. With millions of mobile applications (apps) in multiple genres available for end-users, there is a lack of research on sub-optimal code developers intentionally implement in mobile apps.  

In this study, we examine the occurrence and characteristics of SATD in 15,614 open-source Android apps. Our findings show that even though such apps contain occurrences of SATD, the volume per app (a median of 4) is lower than in non-mobile systems, with most debt categorized as Code Debt. Additionally, we identify typical elements in an app that are prone to intentional sub-optimal implementations. We envision our findings supporting researchers and tool vendors with building tools and techniques to support app developers with app maintenance. 
\end{abstract}


\section{Introduction}
\label{Section:introduction}
The proliferation of mobile devices, in the form of smartphones and tablets, has made it easier and more convenient for society to find information and access services anytime and anywhere. As the leading mobile operating system \cite{MobileOS_market0}, Android plays a crucial role in enabling users to perform various tasks previously unachievable in a mobile environment. Furthermore, Integrated Development Environments, together with frameworks and libraries, make it easier for developers to build feature-rich and interactive mobile applications (apps), as evidenced by the volume and diversity of apps available on the Google Play Store---over 2 million as of September 2022 \cite{GooglePlay_apps}. However, like non-mobile systems, mobile apps are also software systems, and are subject to poor code quality when app developers deviate from established best practices and guidelines \cite{AndroidQualityGuidelines}. For instance, poor code quality of a mobile app can lead to various issues for end-users, such as security vulnerabilities \cite{Shahriar2014ContentProvider}, poor user experience \cite{Caro2018Usability} (including accessibility concerns \cite{Vendome2019Accessibility}), and performance degradation \cite{Linares2015Performance}. Further, app developers face maintainability challenges due to their app's poor code quality \cite{Malavolta2018Maintainability}.

While tool vendors and researchers have produced tools and techniques to assist app developers in locating and correcting bad programming practices (i.e., smells) in their code  \cite{AndroidStudioLint,Habchi2021AndroidSmells,Carvalho2019PresentationSmells}, this is not a one-stop solution to address challenges with app maintainability. Poor code quality is not limited to smells; lack of documentation, partial or missing functionality, lack of testing, and even workarounds contribute to poor code quality and negatively impact the maintainability of the system \cite{Alves2014Ontology,Bavota2016MSR}. Furthermore, there can even be instances of developers knowingly deviating from best practices due to project deadlines or budget constraints \cite{yli2014sources}. A continuous accumulation of these sub-optimal implementations can have serious repercussions, such as increased maintenance costs and an increased risk of project failure. This phenomenon of developers knowingly compromising code and design quality in favor of meeting project deadlines/milestones is known as technical debt \cite{Ernst2021TechnicalDebt}. In this context, there are instances of developers consciously acknowledging the presence of technical debt in their system through documentation, typically taking the form of code comments, known as self-admitted technical debt (SATD) \cite{Potdar2014SATD}.

SATD provides a convenient mechanism to study some of the sub-optimal decisions developers make when implementing a system. Employing SATD, the research community has conducted several studies on these acknowledged implementation compromises, their management, and their correction in open-source and industrial software, various software domains, and programming languages \cite{lenarduzzi2021systematic,Sierra2019Survey,Zampetti2021OpensourceIndustry,Vidoni2021R,Azuma2022Docker}. However, even with all these studies, research examining the occurrence of SATD in mobile apps, specifically Android apps is lacking. Unlike traditional (i.e., non-mobile) systems, mobile apps are highly user-centric and should support various portable devices. Additionally, they are more constrained in size, functionality, security, and resource/energy consumption \cite{Flora2014MobileApps}. For example, in Listing \ref{Listing:example1}, the developer has to include additional code as a workaround to turn off the camera flash for different mobile devices (i.e., Samsung  Galaxy and Samsung Behold II). Also, Samsung Behold II depends on a specific version of the Android SDK (SDK 3). This complex scenario is something that app developers face to support their end-users; something seen less commonly with desktop and web applications.

\begin{lstlisting}[caption={A code snippet example of a developer performing workarounds to support specific smartphones \cite{codeListing}.}, label=Listing:example1, firstnumber = last, escapeinside={(*@}{@*)}]
private void setFlash(Camera.Parameters parameters) {
	/* FIXME: This is a hack to turn the flash off on the Samsung Galaxy. Restrict Behold II check to Cupcake, per Samsung's advice */
	if (Build.MODEL.contains("Behold II") && CameraManager.SDK_INT == 3) {
		parameters.set("flash-value", 1);
	} else {
		parameters.set("flash-value", 2);
	}
    /* This is the standard setting to turn the flash off that all devices should honor. */
    parameters.set("flash-mode", "off");
}
\end{lstlisting}



\subsection{Goal \& Research Questions}
As an exploratory study, the goal of this research is to discover the extent to which Android apps incur SATD and form a high-level understanding of the types and causes of these sub-optimal implementations. Thus, we \textit{analyze the instances of developers documenting sub-optimal implementation in the code as comments}---i.e., the occurrence of SATD comments. We envision findings from this study helping app developers better understand and plan for areas in their apps that are prone to incurring debt. Additionally, our work helps support the development of tools and techniques to assist developers in maintaining their apps. Our study aims at answering the following research questions (RQs):

\vspace{1mm}
\noindent\textbf{\RQA} This RQ reports on how common it is for Android apps to contain SATD and the volume of SATD typically contained within the app. Knowing the extent to which SATD is present in an app will direct us to further research in this area.

\vspace{1mm}
\noindent\textbf{\RQB} Knowing the volume of SATD present in Android apps, this RQ focuses on understanding, at a high-level, the types of SATD present in the app. Hence, we classify SATD comments into one of seven predefined technical debt categories. 

\vspace{1mm}
\noindent\textbf{\RQC} In this RQ, we go one step deeper and examine the SATD comments to understand the typical areas and components in an Android app that are prone to sub-optimal implementation decisions by developers.

\subsection{Contribution}
The main contributions from this work are as follows:
\begin{itemize}
\item This study provides preliminary yet promising findings that expand our awareness of technical debt in Android apps. Through our findings and discussion, the community is made aware of the similarities and differences of technical debt in mobile and non-mobile systems, along with typical elements in an app prone to incurring debt. 
 \item We make our dataset of SATD comments from a large and diverse set of open-source Android apps publicly available for replication and extension purposes \cite{ProjectWebsite}.
\end{itemize}

\section{Related Work}
\label{Section:related_work}
In this section, we provide insight into prior studies in this area. We divide the works into two specific areas -- studies on Android apps and studies on non-mobile systems.

\subsection{Android Apps}
The studies highlighted in this subsection focus on research examining Android apps. While there is plenty of research on the code quality of Android apps (especially smells), we narrow our studies to only those where the authors explicitly mention how their work is related to technical debt.

Verdecchia \cite{Verdecchia2018Android} proposes an approach to identifying architectural technical debt that involves the construction of a reference architecture through a survey of published and grey literature and checking for compliance in reversed engineered apps. However, the authors do not conduct an empirical or case study using their proposed approach. Couto et al. \cite{Couto2020EnergyDebt} present the concept of Energy Debt that occurs in mobile apps due to the existence of energy code smells in its source code. The concepts presented in the study are incorporated into a SonarQube-based tool by Maia et al. \cite{Maia2020EnergyDebt} that calculates the energy debt of Android apps. An analysis of three apps shows that energy debt fluctuates throughout releases. Through a quantitative study of 50 Android apps and developer interviews, Di Gregorio et al. \cite{DiGregorio2022} identify a new type of technical debt, accessibility technical debt, in apps and recommend that developers plan for accessibility early in the project than handling it towards the end of the development phase. Ghari et al. \cite{Ghari2019CASCON} examine the source code of 10 Android apps and show that, among other findings, developers add technical debt to their apps after performing maintenance activities, such as fixing bugs. Habchi et al. \cite{Habchi2019MSR} mine the source code version history of 324 Android apps to identify developers responsible for code smells. The authors show that no single group is responsible for introducing or removing smelly code. The authors also indicate that mobile code smells represent inadvertent technical debt (i.e., debt introduced by oversight than strategically). In \cite{Ramanathan2020Piranha}, Ramanathan et al. describe how their tool efficiently reduces technical debt related to state feature flags on mobile apps, including Android apps, through refactoring.     

\subsection{Non-Mobile Systems}
In this subsection, we focus only on studies related to SATD in non-mobile systems; this includes empirical studies, case studies, and developer surveys. 

An examination of four systems by Potdar and Shihab \cite{Potdar2014SATD} shows that experienced developers are more likely to introduce SATD. The authors also indicate that code complexity does not correlate with the amount of SATD present in the system and that not all SATD is removed in future updates to the code. Wehaibi et al. \cite{Wehaibi2016SANER} study five systems and show that there does not appear to be a relationship between files exhibiting SATD and defects. In a similar study of five systems, Maldonado and Shihab \cite{Maldonado2015} propose the classification of SATD comments into five primary categories-- design debt, defect debt, documentation debt, requirement debt, and test debt. The authors show that most SATD in their dataset is design debt. An empirical study on 159 systems by Bavota and Russo \cite{Bavota2016MSR} shows that systems contain, on average, 51 instances of SATD, with most SATD classified as code debt, followed by requirements and design debt. The authors also indicate that SATD tends to increase during the lifetime of the systems, and, in most cases, the same developer introduces and removes the debt. Finally, the authors do not observe a correlation between the code quality of the source file and the number of SATD instances it contains. Through an analysis of 333 SATD comments, Maipradit et al. \cite{Maipradit2020} introduce an additional category of SATD-- on-hold SATD, representing developers waiting for a specific event/functionality implemented elsewhere. The authors also implement a model to classify on-hold SATD comments. An examination of 2,641 machine learning projects by OBrien et al. \cite{OBrien2022ML} shows that these systems exhibit a high proportion of Requirements debt, followed by Code debt. A study of R packages by Vidoni \cite{Vidoni2021R} shows Code Debt as the most commonly occurring type of SATD. In their study of how SATD is managed in industrial projects, Li et al. \cite{Li2022TSE} examine source code, issue tracker details, and interview developers. The authors observe that most SATD is classified as code/design, while test debt occurs the least. The authors note that industrial projects have more SATD in issues and commit messages compared to open-source projects. In another industry study, through a survey of developers, Zampetti et al. \cite{Zampetti2021OpensourceIndustry} show that industrial developers are less reluctant to document SATD and that SATD is related to functional and maintenance problems. Finally, in a survey of  1,831 developers, Ernst et al. \cite{Ernst2015FSE} report that architectural decisions are the most important source of technical debt and are difficult to address.

\subsection{Summary}
While there are studies on SATD, these are based on non-mobile systems. Looking at the mobile studies, the research community utilizes code smells and other code metrics (e.g., using SonarQube) to study technical debt. In short, we lack the understanding of knowing the common areas in a mobile app which developers deliberately and consciously document their sub-optimal implementation decisions.

\section{Experiment Design}
\label{Section:experiment_design}
In this section, we provide details about the methodology for our study. Figure \ref{Figure:diagram_experiment} shows a high-level overview of our methodology, which we describe in detail below. The dataset we generate in this study is available on our project website for replication and extension purposes \cite{ProjectWebsite}. 

\subsection{Source Dataset}
In this study, we utilize \textit{AndroZooOpen} \cite{AndroZooOpen}, a dataset of 46,522 open-source Android apps spanning multiple categories.
From this source dataset, we attempt to clone each project repository\footnote{The projects were cloned in October 2022.}. However, not all project repositories in \textit{AndroZooOpen} are publicly accessible. Hence, we cloned a total of 37,342 Android app project repositories containing Java files. 

\begin{figure}
 	\centering
 	\includegraphics[trim=0cm 0cm 0cm 0cm,clip,scale=0.75]{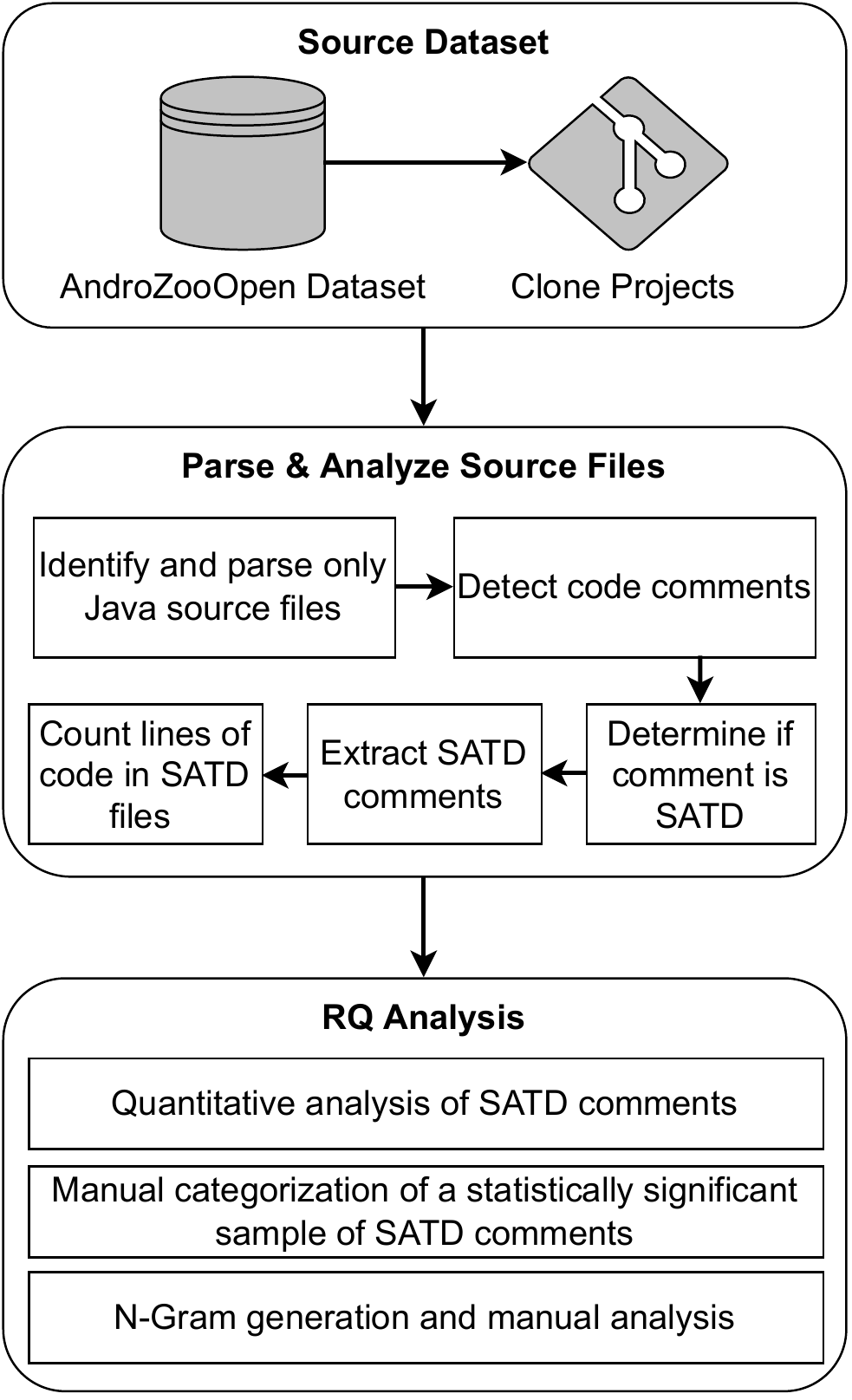}
 	\caption{Overview of our experiment design.}
 	\label{Figure:diagram_experiment}
\end{figure}

\subsection{Parse \& Analyze Source Files}
The next stage of our experiment involves detecting SATD-related comments. It should be noted that our analysis is limited to the files contained in the most recent commit of the cloned repository (i.e., we did not analyze the version history of the files in the project). Below outlines our analysis approach:
\begin{itemize}
    \item First, we recursively analyzed all directories of the cloned projects to identify all Java files by isolating files with a ``.java'' extension (case insensitive). In total, we detected 1,386,942 Java files. Furthermore, we calculated that each project contains a median of 10 Java files.
    \item Next, for each detected Java file, we used \textit{JavaParser}\footnote{\url{https://javaparser.org/}} to construct its corresponding abstract syntax tree (AST). \textit{JavaParser} has been utilized in prior research studies to analyze source code, such as test smell detection \cite{Peruma2020tsDetect}. Through this AST, we can access (i.e., visit) each code comment in the Java file. Furthermore, by using the AST, we ignore the analysis of Java files having compilation issues. In this activity, we capture line, block, and JavaDoc comments.  In total, we detected 14,383,880 comments. From the 37,342 projects that we cloned, only 35,982 projects contained comments in the source code.
    \item Next, we utilized SATD Detector Core \cite{Liu2018SATD} to detect SATD-related comments. This tool takes a comment string as input and then utilizes natural language processing techniques and a machine learning-based binary classifier to classify whether the comment is SATD. We selected this tool as it has been utilized in prior studies \cite{Xavier2020SATDissuetracker,AlOmarSATDBailiff2022,OBrien2022ML}. We report the results of this activity in Section \ref{Section:results}.
    \item Finally, to obtain size-related metrics of the files with SATD comments, we utilize cloc \cite{cloc} to count blank lines, comment lines, and physical lines of source code.  
\end{itemize}

\subsection{RQ Analysis}
\label{Section:experiment_design_rqanalysis}
In this stage of the experiment, we answer our research questions by analyzing the SATD comments. We follow a mixed-methods approach consisting of quantitative and qualitative data analysis. Our quantitative approach includes utilizing well-established statistical measures. Our qualitative activities include manually reviewing a sample set of comments to further understand the technical debt in Android apps.

To this extent, we manually reviewed a statistically significant sample of SATD comments and group related comments into categories. We utilize the six high-level categories proposed by Bavota and Russo \cite{Bavota2016MSR}---Requirements Debt, Design Debt, Code Debt, Defect Debt, Documentation Debt, and Test Debt. Additionally, similar to Azuma et al. \cite{Azuma2022Docker}, we also have an Unclassifiable category. Table \ref{Table:debt-definition} describes the seven categories.

\begin{table}
\centering
\caption{Definitions for the different types of technical debt.}\vspace{-2mm}
\label{Table:debt-definition}
\begin{tabular}{p{0.27\linewidth} p{0.65\linewidth}}
\toprule
\multicolumn{1}{c}{\textbf{Debt Type}} & \multicolumn{1}{c}{\textbf{Description}}                                                \\ \midrule
Requirements Debt                      & Missing or partial implementation of the code.                                           \\ \midrule
Design Debt                            & Structural issues of the codebase, violations of design and object-oriented principles. \\ \midrule
Code Debt &
  Comments that indicate low code quality that makes it challenging to maintain the source code. This includes poor readability, workarounds, dead/redundant code, etc. \\ \midrule
Defect Debt                            & Known defects in the code that are not fixed by the developer.                           \\ \midrule
Documentation Debt                     & Missing or incomplete documentation associated with the code.                            \\\midrule
Test Debt                              & Issues related to the implementation or improvement of existing unit tests.              \\ \midrule
Unclassifiable                         & Comments that reflect technical debt but are vague or incomplete.                        \\ \bottomrule
\end{tabular}\vspace{-3mm}
\end{table}

Further, as this study is on Android apps, we are interested in knowing the common elements in the codebase associated with technical debt. We are especially interested in sub-optimal code related to mobile and Android-specific elements. To achieve this task, we extracted and manually analyzed the top 50 frequently occurring unigrams, bigrams, and trigrams in SATD comments and grouped them into categories. Additionally, we manually searched and reviewed comments containing these terms to provide us with context around their occurrence.

\section{Results}
\label{Section:results}
In this section, we report on the findings of our experiments by answering our RQs. The first RQ examines the volume of SATD present in Android apps. The second RQ explores the types of SATD in the code base of these apps. Finally, the third RQ examines the types of code elements in Android apps most frequently associated with SATD. Due to space constraints, we only show the most frequently occurring types; the complete dataset is available at: \cite{ProjectWebsite}. 

\subsection{\RQA}
\textbf{Motivation:} This RQ aims to understand the extent to which Android app developers knowingly deviate from best practices in building their apps. To this extent, we examine the presence of SATD in the app's source code. From this RQ, we aim to understand the extent to which technical debt occurs in mobile apps compared to non-mobile software systems.

Running SATD Detector Core \cite{Liu2018SATD} on our dataset of 35,982 Android apps that contain comments, we obtain 15,614 (or approximately 43.39\%) apps with SATD comments. Furthermore, as shown in Table \ref{Table:satd}, we observe each app having a median of 4 SATD comments. Furthermore, compared to the total number of comments in an app, SATD comments contribute to, on average, 5.97\% of the comments.

Next, our dataset contains 451,499 SATD-related comments. Looking at the files in the dataset, we observe that a file contains a median of 1 SATD comment and each app contains a median of 2 SATD files. Additionally, Table \ref{Table:freq_satd} shows that most apps (approximately 67\%) contain between 1 to 5 SATD-related comments. Furthermore, we conducted a Spearman correlation test \cite{TaegerStatisticalTesting} to measure the relationship between the number of Java source files and SATD comments in an app. We utilized this nonparametric test as our data does not follow a normal distribution, which we confirmed via a Shapiro-Wilk normality test \cite{TaegerStatisticalTesting}. The Spearman correlation test yielded a statistically significant (i.e., $p-value < 0.05$) correlation of 0.6, equating to a moderate positive correlation. 

Moving on, examining the lines of code in files with and without SATD, we observe that files with SATD have a median of 122 lines of code, while files without SATD have a median of 46 lines of code. To test the statistical significance of this observance, we perform a nonparametric Mann-Whitney-Wilcoxon test on the lines of code for these two groups of files. Our null hypothesis is that there is no difference in the lines of code between files with and without SATD. The results of this calculation yield a statistically significant p-value (i.e., $p-value < 0.05$), causing a rejection of the null hypothesis, meaning that files with and without SATD have different distributions of lines of code.

Finally, as shown in Figure \ref{Figure:CommentType}, looking at the types of SATD comments, we observe that developers frequently utilize Line comments (246,709 instances or 54.64\%) to document technical debt, followed by JavaDoc (182,180 instances), and Block comments (22,610 instances).



\begin{table}
\centering
\caption{Statistical summary of the occurrence of SATD in our dataset.}\vspace{-2mm}
\label{Table:satd}
\begin{tabular}{crrrrr}
\toprule
\textbf{Min.} &
  \multicolumn{1}{c}{\textbf{1st Qu.}} &
  \multicolumn{1}{c}{\textbf{Median}} &
  \multicolumn{1}{c}{\textbf{Mean}} &
  \multicolumn{1}{c}{\textbf{3rd Qu.}} &
  \multicolumn{1}{c}{\textbf{Max}} \\ \hline
\multicolumn{6}{c}{\textit{Count of SATD comments in apps}}  \\
\multicolumn{1}{r}{1}    & 2   & 4   & 28.92  & 12  & 17483  \\ \hline
\multicolumn{6}{c}{\textit{Count of SATD comments in files}} \\
\multicolumn{1}{r}{1}    & 1   & 1   & 2.34   & 2   & 1140   \\ \hline
\multicolumn{6}{c}{\textit{Count of SATD files in apps}}     \\
\multicolumn{1}{r}{1}    & 1   & 2   & 12.05  & 6   & 3860   \\ \bottomrule
\end{tabular}\vspace{-2mm}
\end{table}

\begin{table}
\centering
\caption{Distribution of SATD comments in apps.}\vspace{-2mm}
\label{Table:freq_satd}
\begin{tabular}{@{}rrr@{}}
\toprule
\multicolumn{1}{c}{\textbf{SATD Comments Per App}} & \multicolumn{1}{c}{\textbf{Frequency}} & \multicolumn{1}{c}{\textbf{Percent}} \\ \midrule
1               & 3,808           & 27.83\%          \\
2               & 2,287           & 16.71\%          \\
3               & 1,375           & 10.05\%          \\
4               & 1,008           & 7.37\%           \\
5               & 746             & 5.45\%           \\
\textit{others} & \textit{6,390}  & \textit{46.69\%} \\
\textbf{Total}  & \textbf{15,614} & \textbf{100\%}   \\ \bottomrule
\end{tabular}\vspace{-4mm}
\end{table}

\def\angle{0}
\def\radius{3}
\def\cyclelist{{"orange","yellow","red","blue"}}
\newcount\cyclecount \cyclecount=-1
\newcount\ind \ind=-1
\begin{figure}
\centering
\begin{tikzpicture}[nodes = {font=\sffamily}]
\foreach \percent/\name in {
    54.64/Line,
    40.35/JavaDoc,
    5.01/Block
    } {
      \ifx\percent\empty\else               
        \global\advance\cyclecount by 1     
        \global\advance\ind by 1            
        \ifnum3<\cyclecount                 
          \global\cyclecount=0              
          \global\ind=0                     
        \fi
        \pgfmathparse{\cyclelist[\the\ind]} 
        \edef\color{\pgfmathresult}         
        \draw[fill={\color!50},draw={\color}] (0,0) -- (\angle:\radius)
          arc (\angle:\angle+\percent*3.6:\radius) -- cycle;
        \node at (\angle+0.5*\percent*3.6:0.7*\radius) {\percent\,\%};
        \node[pin=\angle+0.5*\percent*3.6:\name]
          at (\angle+0.5*\percent*3.6:\radius) {};
        \pgfmathparse{\angle+\percent*3.6}  
        \xdef\angle{\pgfmathresult}         
      \fi
    };
\end{tikzpicture}
\caption{Proportion of comment types containing SATD.}\vspace{-3mm}
\label{Figure:CommentType}
\end{figure}
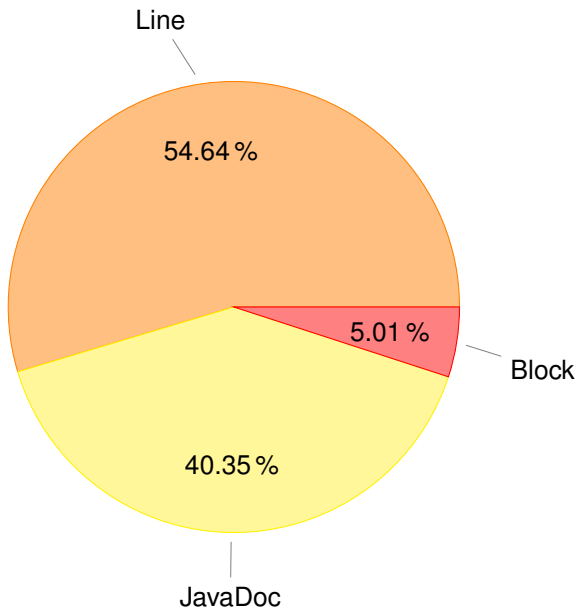

\begin{tcolorbox}[top=0.5pt,bottom=0.5pt,left=1pt,right=1pt]
\textbf{Summary for RQ1.}
Similar to non-mobile system developers, Android app developers make sub-optimal decisions when building their apps. However, the absolute number of SATD comments is lower than in traditional non-mobile systems; a median of four comments. This comparatively low occurrence of SATD is likely due to the fact that apps contain fewer source files.
\end{tcolorbox}

\subsection{\RQB}
\textbf{Motivation:} The prior RQ shows that mobile systems, specifically Android apps, are also prone to technical debt in similar amounts as in non-mobile systems. In this RQ, we go deeper into our detected SATD comments by grouping these comments into seven predefined categories. The categories represent the types of documented technical debt commonly occurring in software systems  (as defined in Section \ref{Section:experiment_design_rqanalysis}). Through this RQ, we gain an understanding of the types of documented technical debt frequently occurring in Android apps. We manually reviewed and annotated a random, statistically significant sample of 386 SATD comments. This sample represents a 95\% confidence level and a 5\% confidence interval. All authors reviewed and annotated each comment, and conflicts were resolved through discussion. Table \ref{Table:type-count}, shows the distribution of these categories---Code Debt occurs the most (35.23\%), followed by Design Debt (23.58\%) and Requirement Debt (19.69\%). We discuss our findings below.

\subsubsection*{\textbf{\textcolor{purple}{\faTag\hspace{2.mm}Code Debt}}} 
Code Debt contributes to the highest amount of SATD instances in our sample. In this category, similar to \cite{Bavota2016MSR}, we include comments noting workarounds developers implement. Workarounds represent compromises developers make to the code's quality to achieve a specific goal which negatively impact maintenance. These workarounds are usually in response to fixing a bug (e.g., ``work around nested unclipped SaveLayer bug''), which can include issues related to APIs the app uses, including Android SDK APIs (e.g., ``\textit{This is to work around a bug in DatePickerDialog where it doesn't display a title showing}''), and specific devices (e.g., ``\textit{Also workaround for bug on Nexus 6...}''). Other observations include:
\begin{itemize}
    \item the need to check the value of an identifier (e.g., ``\textit{TODO: check ret val!}'')
    \item needing to rename an identifier name to something more appropriate (e.g., `\textit{`TODO: Rename parameter arguments, choose names that match}'')
    \item the need to change an identifiers datatype (e.g., ``\textit{TODO: Use strings for id's too}'')
    \item creating placeholders for missing code (e.g., ``\textit{TODO add your handling code here}'')
    \item determining if a method should be called in a particular location (e.g., ``\textit{FIXME: do we need release() here?}'')
\end{itemize}

\subsubsection*{\textbf{\textcolor{purple}{\faTag\hspace{2.mm}Design Debt}}}
The comments within this category relate to shortcomings or workarounds in the structure or design of the app's codebase. This includes the need to move (i.e., extract) code to more appropriate locations (e.g., ``\textit{TODO: Factor out this Glide.get() call}''), which can even include the creation of new classes (e.g., ``\textit{TODO move this out to its own class}''). We also see developers explicitly mention that their code should be refactored (e.g., ``\textit{That's a larger refactoring we'll save for another day}'') and object-oriented principles, such as encapsulation and abstraction (e.g., ``\textit{TODO: CentralSurfaces should be encapsulated behind a Controller}'' and ``\textit{NOTE: Ideally, we would abstract away the details of what identifies a network of a specific}").

\subsubsection*{\textbf{\textcolor{purple}{\faTag\hspace{2.mm}Requirement Debt}}}
This is the third-highest category that contains comments in which developers document functionality that is not implemented or incomplete. These requirements associated with the comments are related to presentation layer elements (e.g., ``\textit{TODO Hide the loading indicator}''), business logic (e.g., `\textit{`TODO Add more comparision later}''), and the data access layer (e.g., ``\textit{TODO delete table data}''). Additionally, developers also document the need to handle future releases of the Android operating systems, or backward compatibility for older versions of the operating systems (e.g. ``\textit{TODO: We'll have that on Android 2.2}'').

\subsubsection*{\textbf{\textcolor{purple}{\faTag\hspace{2.mm}Defect Debt}}}
This category includes comments in which developer document known defects or deficiencies with their apps. Our analysis shows three types of defect debt comments:
\begin{itemize}
    \item functional defects - these defects are usually associated with elements in the presentation layer, such as the screen (e.g., ``\textit{too high, fix height}'') and hardware related (e.g., ``\textit{``XXX: Disables take picture button}'') 
    \item non-functional defects - these defects usually deal with the app's performance and missing error handling (e.g., ``\textit{XXX: This might potentially cause stalls in the main}'' and ``\textit{TODO: handle exception}'')
    \item generic defects - these are defects that lack a description of the problem (e.g., ``\textit{FIXME, this is wrong}'')
\end{itemize}

\subsubsection*{\textbf{\textcolor{purple}{\faTag\hspace{2.mm}Documentation Debt}}}
Our sample only shows 11 instances of Documentation Debt. While most of these comments are JavaDoc, we also see a few line comments. Examining the comments, we observe that developers do not document the purpose/behavior of code and use comments as reminders to explain the behavior in the future. For example, the comment ``\textit{TODO: document}'' represents missing details. We also encounter incomplete code documentation instances, such as ``\textit{We aren't going to go into detail about how this method works, but feel free to explore!}''. Finally, we also see redundant comments, such as  ``\textit{Now upgrade should work fine}".

\subsubsection*{\textbf{\textcolor{purple}{\faTag\hspace{2.mm}Test Debt}}}
Prior work on mobile app testing by Pecorelli et al. \cite{Pecorelli2020ICPC} shows that Android apps are poorly tested, with most app developers not writing tests for their apps, and even for those apps having tests, they have a median of just two tests. This is consistent with the low occurrence of Test Debt in our dataset. Analyzing the five comments in our sample shows developers documenting the need to create/run tests for the implemented functionality (e.g., ``\textit{TODO Test SharedPreferences}''). Additionally, Android Studio includes example unit tests when a project is created. These files, named `ExampleInstrumentedTest.java' and `ExampleUnitTest.java', contain simplistic examples and should be removed \cite{Peruma2019TestSmells}. Unfortunately, Android Studio does not include comments in the file informing the developer about removing them from the project. Looking at these files in our dataset, we observed instances where developers included actual test cases in these files, including SATD-related comments (e.g., ``TODO: Work on the test tmr''). This pattern of sub-optimal testing negatively impacts correctness and maintainability.

\subsubsection*{\textbf{\textcolor{purple}{\faTag\hspace{2.mm}Unclassifiable}}}
Comments in this category are considered SATD since they contain known SATD keywords (i.e., `\textit{TODO}', `\textit{FIXME}', and `\textit{XXX}'). However, these comments are incomplete or vague; examples include: ``\textit{XXX ???}'' and ``\textit{TODO it's bad}''. Developers include these comments for a reason, but since they are not descriptive, we cannot classify them. Furthermore, these ambiguous comments will almost surely hamper maintenance activities, especially on projects staffed by multiple developers.

\begin{table}
\centering
\caption{Distribution of the different types of technical debt in a statistically significant sample of SATD comments.}\vspace{-2mm}
\label{Table:type-count}
\begin{tabular}{@{}lrr@{}}
\toprule
\multicolumn{1}{c}{\multirow{2}{*}{\textbf{Technical Debt Type}}} &
  \multicolumn{1}{c}{\multirow{2}{*}{\textbf{Frequency}}} &
  \multicolumn{1}{c}{\multirow{2}{*}{\textbf{Percentage}}} \\
\multicolumn{1}{c}{}    & \multicolumn{1}{c}{}  & \multicolumn{1}{c}{}    \\ \midrule
Code Debt               & 136                   & 35.23\%                 \\
Design Debt             & 91                    & 23.58\%                 \\
Requirement Debt        & 77                    & 19.95\%                 \\
Defect Debt             & 40                    & 10.36\%                 \\
Unclassifiable          & 26                    & 6.74\%                  \\
Documentation Debt      & 11                    & 2.85\%                  \\
Test Debt               & 5                     & 1.30\%                  \\
\textit{\textbf{Total}} & \textit{\textbf{386}} & \textit{\textbf{100\%}} \\ \bottomrule
\end{tabular}\vspace{-4mm}
\end{table}

\begin{tcolorbox}[top=0.5pt,bottom=0.5pt,left=1pt,right=1pt]
\textbf{Summary for RQ2.}
Code debt contributes the greatest number of SATD instances that Android developers injected into their apps, implying that developers frequently compromise their code quality via workarounds to achieve a desired result. Developers also make sub-optimal design decisions when building their apps and build apps with missing or incomplete requirements.
\end{tcolorbox}

\subsection{\RQC}
\textbf{Motivation:} While RQ1 and RQ2 focus on the volume and types of SATD occurring in Android apps, this RQ investigates the code elements (especially mobile and Android-specific features) most commonly associated with technical debt. This RQ provides insight into areas of concern that Android app developers and project teams can focus on during the implementation and maintenance of apps. To understand this we extracted common n-grams---unigrams, bigrams, and trigrams---from the collected SATD comments. Unigrams are single words; bigrams are pairs of consecutive words occurring in a sentence; trigrams are three consecutive words. Bigrams and trigrams provide more context around the words but can lead to more project-specific terminology and noise.  We manually examined the top 50 n-grams from each set (paying special attention to mobile or Android-specific terminology) and grouped related terms into categories that the authors agreed upon. We also examined the comments containing these terms to gain more insight into the actual technical debt issues. 

Our analysis of these terms yielded four high-level categories---Android SDK API, General UI, General Programming, and Hardware. Some of these categories contain subcategories. Below we elaborate on each category and include illustrative examples.

\subsubsection*{\textbf{\textcolor{purple}{\faTag\hspace{2.mm}Android SDK API}}} Within this category, we encounter comments involving developers documenting sub-optimal code related to using Android-specific features, such as UI-related APIs/components and non-UI-specific APIs the app utilizes. 

Under the \textbf{{\textcolor{violet!80}{Android UI API}}} subcategory, we encounter terms such as `Activity', `View', `Fragment', `Listener', `ActionBar', and `BoundingBox'. Examining the Activity-related comments, we notice developers documenting bugs/issues related to the activity (e.g., ``\textit{TODO: We need to fix this case...}'') and updating/implementing an activity (``\textit{TODO Change to activity to be able to search}'' and ``\textit{TODO (1) Use Android Studio's Activity wizard to create a new Activity...}''). Examining the Fragment comments, we observe developers indicating the need to replace/use specific Fragment types (e.g., ``\textit{TODO: use dialog fragment}'') or optimize existing Fragments (e.g., ``\textit{TODO beautify the fragment}'' and ``\textit{TODO: Make the create fragment layout scrollable}''). Looking at the View comments, we encounter comments about implementing or updating specific Views (e.g., ``\textit{TODO Add a View to the layout with a width of match\_parent and a height of 1dp}''). 

In the \textbf{{\textcolor{violet!80}{Android non-UI API}}} subcategory, some frequent terms we encounter include `Context', `Loader', `ContentProvider', and `AndroidManifest'. These terms are known components in the Android SDK, which developers utilize when building apps.  We observe developers documenting the need for using a ContentProvider for existing features (e.g., ``\textit{TODO Set up a content provider interface to abstract contacts from phone!}''). Looking at AndroidManifest comments, we observe developers documenting the need to add or remove entries from this file due to functional and quality attribute changes (e.g., ``\textit{TODO: 01. Add Fingerprint Feature + Permission in AndroidManifest.xml}''). Additionally, we encounter the terms `deprecated\_api' and `api\_level', which developers utilize to indicate if their code is compatible/incompatible with a specific Android API version (e.g., ``\textit{TODO: use API level 24 or above to correct this}'' and ``\textit{TODO: delete these deprecated method calls once we support only API 23 and higher}'').

\subsubsection*{\textbf{\textcolor{purple}{\faTag\hspace{2.mm}General UI}}} This category contains general user interface terms such as `button', `button\_clicked', `pixels', `margin', `form', `color', and `position'. When we examine the comments, we observe that most of the comments associated with these terms are related to implementing new or missing functionality related to the apps' user interface. For instance, the following comments were added by developers relating to a button and color in their apps respectively, ``\textit{TODO: Log the button press as an analytics event}'' and ``\textit{TODO: Add support for border color and types}''.  

\subsubsection*{\textbf{\textcolor{purple}{\faTag\hspace{2.mm}General Programming}}} In this category, we encounter terms related to general programming concepts or activities. We grouped these terms into the following subcategories: Refactoring, Date/Time, Storage, External Resources, Error Handling, and Security.

Within the \textbf{\textcolor{violet!80}{Refactoring}} subcategory, we encounter terms such as `update\_argument\_type' and `rename\_change', showing that app developers document the need to improve the quality of their code from simple identifier renaming to more complex design changes. For example: ``\textit{TODO: Rename and change types and number of parameters}''. 

Under the \textbf{\textcolor{violet!80}{Date/Time}} subcategory are terms like `time\_zone' and `utc\_date'. Since most mobile apps cater to users worldwide, apps must consider a user's location and timezone when performing date/time-related functionality. Looking at the comments, we encounter instances of developers documenting the need to handle timezones (e.g., ``\textit{TODO take care of time zone?}'') or optimizing existing timezone conversions (e.g., ``\textit{TODO: clumsy: implicit conversion from UTC to YYYYMMDDHHMMSS in begin.setTimestamp}'').

The \textbf{\textcolor{violet!80}{Storage}} subcategory represents technical debt related to data storage by the app and includes terms such as `data', `json', `distinct\_rows', and `database\_contains\_tables'. The comments show the need to implement functionality to convert data to a specific format and also parse data in specific formats like JSON (e.g.,  ``\textit{TODO: Send the queue in JSON format}'' and ``\textit{TODO: parse json and populate the user data}''). Database-related problems include improving the design of the data access layer code (e.g., ``\textit{Refactor into Database interface}''), which can also lead to performance improvements (e.g., ``\textit{TODO: database operations should be done on separate thread}'') and better error handling (e.g., ``\textit{TODO: Handle database error}''). Additionally, we observe comments about updating existing functionality to utilize a database to store app data (e.g., ``\textit{TODO: get from a json/csv files/ database etc. But for now hard coded}''). 

Within the \textbf{\textcolor{violet!80}{External Resources}} subcategory, we encounter terms like `url', `link', `server', ` connection\_available', and `data\_request\_server'. The technical debt in this category is related to the app communicating/connecting with external resources, which is usually a server. This includes retrieving data from a server (e.g., ``\textit{TODO get image from server if exists}''), checking the availability of external resources, and taking necessary action on connection failures or resource unavailability (e.g., ``\textit{TODO When Internet Connection un-available}''). We also encounter instances where developers utilize placeholder/test URLs and document the need to replace them (e.g., ``\textit{TODO: switch to PROD URL on release}'').
   
The \textbf{\textcolor{violet!80}{Error Handling}} subcategory involves developers documenting known issues and the need for better error handling. Terms in this subcategory include `bug\_start', `bug\_end', `throws\_ioexception', and `data\_throws\_jsonexception'. Looking at the comments, we observe developers either acknowledging that some code statements are susceptible to runtime exceptions  (e.g., ``\textit{TODO: stop() throws an exception if you haven't fed it any data.  Keep track}'') or the need for further testing to check how the app handles unforeseen errors (e.g., ``\textit{TODO: what if this throws an exception?}''). Additionally, developers also document known bugs in the app (e.g., ``\textit{TODO there are bug in rotation mainly...}'').

Under the \textbf{\textcolor{violet!80}{Security}} subcategory, we observe comments about improving the security of the app (e.g.,  `\textit{`TODO should apply better security policy!}''), potential vulnerabilities (``\textit{TODO: FIXME: This is a potential security problem!}''), and non-optimal workarounds that address security concerns (e.g., ``\textit{Hacky solution as part of fixing a security bug; ignore}''). We also observe comments around permissions, such as checking if permissions have been denied (e.g., ``\textit{TODO Add Permission check}'') and informing users about the status of permissions (e.g., ``\textit{Todo ask for SEND Message permission}'').

\subsubsection*{\textbf{\textcolor{purple}{\faTag\hspace{2.mm}Hardware}}} This category includes terms like `camera' and `device' and corresponds to code that integrates with specific mobile device components. For example, developers implementing workarounds to support specific devices (e.g., ``\textit{the front-facing camera, its just a hack not all device camera apps support these extras}''), or the need to incorporate or update features related to specific hardware components (e.g., ``\textit{TODO : add usb device attached intent}'' and ``\textit{TODO: For now, assume the device supports LTE}'').

\begin{tcolorbox}[top=0.5pt,bottom=0.5pt,left=1pt,right=1pt]
\textbf{Summary for RQ3.}
As evident from the comments in the codebase, Android app developers introduce technical debt associated with four categories: Android API/Features, General UI, General Programming, and Hardware. Furthermore, some of these categories are composed of multiple subcategories that provide a more granular view of potential sub-optimal code in Android apps. 
\end{tcolorbox}

\section{Discussion}
\label{Section:discussion}
As an exploratory study, our research aims to understand the reasons for SATD in Android apps. Our analysis of code comments shows that developers take shortcuts or make sub-optimal decisions in building apps, the common types of debt the project accumulates, and the specific components and features associated with technical debt. While this study expands the body of knowledge in mobile app development and maintenance, our findings also suggest further research.  In this section, we discuss how our work complements and aligns with existing research on Android apps and its implications as a series of takeaways.

While RQ1 shows that Android apps, like traditional, non-mobile systems, are not exempt from technical debt, we also see some differences with existing literature. At the time of conducting this research, ours is the only study examining SATD in Android apps; hence, our comparison is against studies on non-mobile systems. For instance, prior studies on non-mobile systems show that SATD comments contribute to, on average, between 15.01\% to 22.51\% of comments per system \cite{Maldonado2015, Wehaibi2016SANER}. In contrast, our data shows SATD comments contribute, on average, to 5.97\% of the comments in an Android app. This difference can be attributed to the size of the mobile and non-mobile systems; Android apps are smaller and are often developed by small teams \cite{Tufano2015ICSE}. Non-mobile systems may have hundreds or thousands of classes, while the number of classes in mobile apps is typically in the double digits. The apps in our dataset have a median of 10 files/classes (similar to what has been reported elsewhere  \cite{Mojica2014AppSize}).

Our RQ2 findings also show some similarities and contrasts with non-mobile systems. Similar to Bavota et al. \cite{Bavota2016MSR}, Code debt occurs most often in our dataset. However, we have contrasting rankings for the other categories. The low occurrence of requirement and defect debt might be attributed to mobile apps being small in size and having limited functionality. But this observation invites future research.

The findings from RQ3 align with prior work on mobile systems. For instance, work by Carvalho et al. \cite{Carvalho2019PresentationSmells} on the presentation layer identifies code smells involving components such as Activities, Fragments, and Listeners, which developers also mention in SATD comments. Furthermore, Content Providers, which are associated with leakage vulnerabilities \cite{Shahriar2014ContentProvider}, are also present in our findings. This opens up an interesting avenue of research for the community to examine the extent to which such debt items can lead to app vulnerabilities. Research also shows that Android permissions are a cause for concern \cite{Scoccia2019Permission}, which is highlighted in our findings. Our findings about app developers documenting the need to refactor their code, such as renaming identifiers, are reflected in the existing literature that shows app developers refactor their code to improve code comprehension, among other reasons \cite{Peruma2019Refactoring}. Additionally, our identification of sub-optimal Error Handling in Android apps aligns with research showing that exception handling is a problematic area in non-mobile systems \cite{Digkas2017TDapache}. 

Below, we discuss how the findings from our RQs support the community through a series of takeaways.

\vspace{1mm}
\noindent\faThumbTack\hspace{0.08cm}\textbf{Takeaway 1 - \textit{Integration of technical debt detection tools into the developer workflow.}} There are a number of tools and techniques to detect the presence of SATD in code \cite{Sierra2019Survey}. IDE vendors and developers should utilize these mechanisms in the development workflow. For instance, developers can integrate these tools into the build process to receive notifications of the presence of technical debt when code is committed. Additionally, IDE vendors integrating these tools into their products can provide developers with real-time notifications of sub-optimal code. Note, however, that a recent study has cast doubt on the efficacy of several existing tools \cite{lefever2021}. 

\vspace{1mm}
\noindent\faThumbTack\hspace{0.08cm}\textbf{Takeaway 2 - \textit{Expand research into mobile technical debt.}} While the research community has made strides in mobile quality research, such as code and test smells \cite{Hamdi2021Smells,Peruma2019TestSmells}, our work provides an opportunity for further research into implementing and maintaining mobile apps. For instance, our findings from RQ3 highlight specific Android APIs usually associated with  debt. These findings provide the community with an avenue to build or enhance code quality tools to support app developers, such as expanding the catalog of Android code smells. Additionally, similar studies on iOS apps can provide insight into unique iOS features associated with debt. 

\vspace{1mm}
\noindent\faThumbTack\hspace{0.08cm}\textbf{Takeaway 3 - \textit{Expand techniques and tools to refactor mobile app technical debt.}} Past studies show the co-occurrence of refactoring actions with SATD removal \cite{Iammarino2021empirical}, with developers performing the refactoring to remove specific  debt items \cite{Peruma2022MSR}. With Android app developers making sub-optimal mobile-specific decisions when building their apps, there exists an opportunity to develop refactoring operations geared toward mobile app code, specifically UI/presentation layer code. This also includes improving the accuracy of existing refactoring recommendation tools by considering the text of the comments and their related code statements.

\vspace{1mm}
\noindent\faThumbTack\hspace{0.08cm}\textbf{Takeaway 4 - \textit{Complement tool use with code reviews.}} Developers should not treat code quality tools as a one-stop solution. For instance, while tools exist to detect smells, including Android-specific smells, it should be noted that not all technical debt is due to the presence of smells. To this extent, project teams should complement their use of tools by conducting frequent code reviews. Furthermore, while it might not be feasible to repay the debt in all situations, the review process can catch instances of poorly composed SATD comments, such as vague or incomplete comments.


\section{Threats To Validity}
\label{Section:threats}
Even though our dataset is restricted to open-source Android apps, the large volume of apps (15k+) contained in the dataset provides a diverse and representative sample. Additionally, since the apps we analyzed are implemented in Java, it helps us to compare results reported in prior literature that analyzed non-mobile Java systems. Further, some apps in this dataset are available in app stores for end-users to install on their device. However there is a threat to external validity in that we can not assume these results apply to languages other than Java. 

Another threat to external validity comes from the fact that we are only examining \textit{self-admitted} technical debt. This means that developers are conscious of the debt that they have introduced or are saddled with. But other forms of debt may be less obvious to them, such as architectural debt \cite{xiao2022} or energy debt \cite{Couto2020EnergyDebt}. For this reason, we may not be able to generalize these results to all forms of technical debt.

The tool we utilized to detect SATD comments has been utilized in similar studies. However, there is a threat of false positives in the dataset; other tools like Pilot \cite{SalleTechDebt2022} and DebtHunter \cite{SalaEASE2021} might provide different results. That said, our RQ2 and RQ3 approaches involved the manual analysis of comments, ensuring that we analyzed only SATD comments. Additionally, our RQ2 and RQ3 approaches also involve peer-reviews of author annotations as means to avoid bias. Further, even though we utilized seven predefined categories in RQ2, these categories are common to other SATD studies.

\section{Conclusion \& Future Work}
\label{Section:conclusion}
The ease of Android app development has resulted in the proliferation of apps that provide end-users access to information and services on multiple mobile devices. However, like traditional systems, mobile apps are also subject to poor coding practices that hamper maintenance. In this exploratory empirical study on over 15k open-source Android apps, we examined how developers documenting sub-optimal implementation decisions---SATD comments.  Our findings show that even though Android apps are not exempt from technical debt, the volume of SADT they exhibit is lower than non-mobile systems. Our findings also show that most debt falls under the Code debt category, of which most are implementation of workarounds. We also show that technical debt is not only related to general programming or design concepts but also due to shortcomings developers take when implementing code that utilizes Android APIs. Additionally, we see developers making sub-optimal decisions in implementing their app's UI.

Our future work in this area includes examining the repayment and survival of SADT in Android apps, by analyzing the version history of source files. These findings will help us better understand the similarities and differences in how developers implement and maintain apps compared to non-mobile systems and give researchers and vendors more insight into how to better support app developers.

\bibliographystyle{ieeetr}
\bibliography{main}
\end{document}